\documentclass[a4paper,epsfig,12pt]{article}
\usepackage[pctex32]{graphics}

\begin{document}
\topmargin 0pt \oddsidemargin 0mm

\renewcommand{\thefootnote}{\fnsymbol{footnote}}
\begin{titlepage}
\begin{flushright}
INJE-TP-03-07\\
hep-th/0307045
\end{flushright}

\vspace{5mm}
\begin{center}
{\Large \bf Holographic temperature bound in the slow-roll
inflation } \vspace{12mm}

{\large  Yun Soo Myung\footnote{e-mail
 address: ysmyung@physics.inje.ac.kr}}
 \\
\vspace{10mm} {\em  Relativity Research Center and School of
Computer Aided Science, Inje University Gimhae 621-749, Korea}
\end{center}

\vspace{5mm} \centerline{{\bf{Abstract}}}
 \vspace{5mm}
We investigate  the relationship between the holographic
temperature bound and the slow-roll inflation. For this purpose we
introduce the  holographic temperature  bound for a radiation
matter :$T \ge T_{\rm H}$. Here $T_{\rm H}$ is the Hubble
temperature which arises from the cosmological holographic
description of for a radiation-dominated universe. For the
quasi-de Sitter phase of slow-roll inflation, we find that  the
holographic temperature bound of $T_{\rm GH} \ge T_{\rm H}$ is
guaranteed with the Gibbons-Hawking temperature $T_{\rm GH}$. When
$T_{\rm GH}= T_{\rm H}$, inflation ends.

\end{titlepage}

\newpage
\renewcommand{\thefootnote}{\arabic{footnote}}
\setcounter{footnote}{0} \setcounter{page}{2}

The inflation turned out to be a successful tool to resolve the
problems of the  hot big bang model~\cite{Infl}. Thanks to the
recent observations of the cosmic microwave background
anisotropies and large scale structure galaxy surveys, it has
become widely accepted by the cosmology community~\cite{JGB}. The
idea of inflation is based on the very early universe dominance of
vacuum energy density of a hypothetical scalar field, the
inflaton. This produces the quasi-de Sitter spacetime~\cite{Hogan}
and during the slow-roll period, the equation of state can be
approximated by the vacuum state as $p\approx -\rho$. After that
there must exist a strong non-adiabatic and out-of-equilibrium
phase called reheating to produce a large increase of the entropy.
However we don't know exactly how inflation started.

On the other hand,  the implications of the holographic principle
for cosmology have been investigated in the
literature~\cite{Hooft,Beke,FS,Hubb,Bous}.
 Verlinde proposed the cosmological  holographic bound  Eq.(\ref{2eq7}) in
a radiation-dominated phase by introducing three
entropies~\cite{Verl}. As an  example, such a radiation-dominated
phase is provided by a conformal field theory (CFT) with a large
central charge which is dual to the AdS-black hole~\cite{SV}. In
this case it appeared  an interesting relationship between the
Friedmann equation governing the cosmological evolution and the
square root form of entropy-energy relation, called Cardy-Verlinde
formula~\cite{Cardy}. Although the Friedmann equation has the
geometric origin and  the Cardy-Verlinde formula is designed only
for the matter content, it  suggested  that both  may arise from a
single  fundamental theory.

In this work we will explore the implications of the holographic
principle for  describing the slow-roll inflation. It is not easy
to obtain the holographic bounds for this period, compared with a
radiation-dominated universe~\cite{Myung}. However, considering a
quasi-de Sitter phase of slow-roll inflation leads to the
holographic temperature bound. This work will provide a  solution
to the question of how   the holographic principle is useful to
describe inflation.

Let us start an $(n+1)$-dimensional Friedmann-Robertson-Walker
(FRW) metric
\begin{equation}
\label{2eq1} ds^2 =-d\tau^2 +R(\tau)^2 \Big[ \frac{dr^2}{1-kr^2} +r^2 d\Omega^2_{n-1} \Big],
\end{equation}
where $R$ is the  scale factor of the universe and
$d\Omega^2_{n-1}$ denotes the line element of an
$(n-1)$-dimensional unit sphere. Here $k=-1,~0,~1$ represent that
the universe  is open, flat, closed, respectively. A cosmological
evolution is determined by the two Friedmann equations
\begin{eqnarray}
\label{2eq2}
 && H^2 =\frac{16\pi G_{n+1}}{n(n-1)}\frac{E}{V}
-\frac{k}{R^2}, \nonumber \\
&& \dot H =-\frac{8\pi G_{n+1}}{n-1}\left (\frac{E}{V} +p\right)
    +\frac{k}{R^2},
\end{eqnarray}
where $H$ represents the Hubble parameter with the definition
$H=\dot R/R$ and the overdot stands for  derivative with respect
to the cosmic time $\tau$,  $E$ is the total energy of matter
filling the universe, and $p$ is its pressure. $V$ is the volume
of the universe, $V=R^n \Sigma^n_k$ with $\Sigma^n_k$ being the
volume of an $n$-dimensional space with $k$, and $G_{n+1}$ is the
Newton constant in ($n+1$) dimensions. Here we assume the equation
of state:
 $p=\omega \rho,~ \rho=E/V$.
Before we proceed, we introduce  three  entropies
 for a holographic description of a  radiation-dominated universe~\cite{Verl}:
\begin{eqnarray}
\label{2eq3}
 {\rm Bekenstein-Verlinde\ entropy}:&& S_{\rm
 BV}=\frac{2\pi}{n}ER,
   \nonumber \\
 {\rm Bekenstein-Hawking\ entropy}:&& S_{\rm
 BH}=(n-1)\frac{V}{4G_{n+1}R},
    \nonumber \\
  {\rm Hubble\ entropy}:&& S_{\rm H}=(n-1)\frac{HV}{4G_{n+1}}.
\end{eqnarray}
 We  define a quantity $E_{\rm BH}$ which
corresponds to energy needed to form a universe-size black hole: $
S_{\rm BH}=(n-1)V/4G_{n+1}R \equiv 2\pi E_{\rm BH} R/n $. The
Friedmann equations (\ref{2eq2}) can be further cast to the
cosmological Cardy-Verlinde formula and cosmological Smarr formula
respectively
\begin{eqnarray}
\label{2eq4}
 && S_{\rm H}=\frac{2\pi R}{n}\sqrt{E_{\rm
BH}(2E-kE_{\rm
BH})}, \nonumber \\
&& kE_{\rm BH}=n(E+pV -T_{\rm H} S_{\rm H}),
\end{eqnarray}
where the Hubble temperature ($T_{\rm H}$) is given by
\begin{equation}
\label{2eq5}
 T_{\rm H}=-\frac{\dot H}{ 2\pi H}
 \end{equation}
 as the minimum
temperature during the strongly gravitating phase of $HR \ge
\sqrt{2-k}$.  Eq.(\ref{2eq4}) corresponds to another
representation of the Friedmann equations expressed in terms of
holographic quantities.

On the matter-side, the entropy of radiation and its Casimir
energy can be described by the Cardy-Verlinde formula and the
Smarr formula, respectively
\begin{eqnarray}
\label{2eq6} && S =\frac{2\pi R}{n}\sqrt{E_c(2E-E_c)},
  \nonumber \\
&& E_c=n(E+pV -T S).
\end{eqnarray}
 The first denotes  the entropy-energy relation, where $S$ is the entropy of a CFT-like radiation living on
an $n$-dimensional sphere with radius $R$ and  $E$ is the total
energy of the CFT. The second represents the relation between  a
non-extensive part of the total energy (Casimir energy) and
thermodynamic quantities.  Here $E_c$ and $T$ stand for the
Casimir energy of the system and the temperature of radiation with
$\omega=1/n$. We note again that the above equations correspond to
thermodynamic relations for the CFT-matter which are originally
independent of the geometric Friedmann equations.  Suppose that
the entropy of radiation in the FRW universe can be described by
the Cardy-Verlinde formula. Then comparing (\ref{2eq4}) with
(\ref{2eq6}), one  finds that if $E_{\rm BH}=E_c$, then $S_{\rm
H}=S$ and $T_{\rm H}=T$. For a $k=1$ closed radiation-dominated
universe, a bound on the Casimir energy ($E_c \le E_{\rm BH}$)
leads to the Hubble bounds for entropy and temperature
~\cite{Verl}
 \begin{equation}
 \label{2eq7}
 S \le S_{\rm H},~~ T \ge T_{\rm H}, ~~{\rm for}~ HR \ge 1
 \end{equation}
which shows  inequalities between geometric and matter quantities.
The Hubble entropy bound can be saturated by the entropy of a
radiation-matter filling  the universe when its Casimir energy
$E_c$ is
 enough to form a universe-size black hole.
If this happens, equations (\ref{2eq4}) and (\ref{2eq6}) coincide.
  This implies that the first Friedmann equation
somehow knows the entropy formula  for a radiation-matter filling
the universe. As an  example, we consider a moving brane universe
in the background of the 5D Schwarzschild-AdS black hole. Savonije
and Verlinde~\cite{SV} found that when this  brane crosses the
black hole horizon, the Hubble entropy bound  is saturated by the
entropy of black hole(=the entropy of the CFT-radiation). At this
moment ($T_{\rm H},E_{\rm BH}$) are identical with ($T,E_c$) of
the CFT-matter dual to the AdS black hole respectively. For a
radiation-dominated universe with a positive cosmological
constant, the  holographic bound was discussed in~\cite{CM1}.

 For arbitrary
$k$,  the Hubble  bounds for a radiation-dominated universe are
still valid ~\cite{Youm2}
 \begin{equation}
 \label{2eq8}
 S \le S_{\rm H},~~T \ge T_{\rm H},~~{\rm for}~ HR \ge  \sqrt{2-k}.
 \end{equation}
On the other hand,  for the general equation of state of $p=\omega
\rho$, the entropy-energy relation no loner  coincide with the
first Friedmann equation and the conjectured bound on the Casimir
energy  does not leads to the Hubble entropy bound. But it was
argued that even for $\omega \not= 1/n$ including $\omega=-1$ for
a cosmological constant $\Lambda$, the Hubble temperature bound
($T \ge T_{\rm H}$) is still satisfied~\cite{Youm1}. We wish to
test here whether or not this argument is correct for the
inflation.

 For this purpose, we adopt a model of
primordial inflation based on the quasi-de Sitter space and FRW
space~\cite{Hogan}. In what follows we work with the
(3+1)-dimensional flat FRW slicing of de Sitter spacetime, because
 this maps directly onto the FRW spacetime of  the
 post inflationary universe. The line element which covers half
 of the full de Sitter solution is given by
\begin{equation}
\label{3eq1} ds^2_{FRW-dS}=-d\tau^2 + \exp[2H\tau]\Big(dr^2 +r^2
d\Omega_2^2 \Big).
\end{equation}
Another slicing of de Sitter spacetime is given by the static
coordinates\footnote{ The coordinates of $\tau,r$ and $t,\tilde r$
are related by the transformations $r=e^{H t} \frac{\tilde
r}{\sqrt{ 1-H^2\tilde r^2}}, \tau= t +\frac{1}{2H} \ln[1-H^2
\tilde r^2]$~\cite{FKo}.}
\begin{equation}
\label{3eq2} ds^2_{s-dS}=-\Big(1-H^2 \tilde r^2\Big)dt^2  +
\Big(1-H^2 \tilde r^2 \Big)^{-1}d \tilde r^2 + \tilde
r^2d\Omega^2_2,
\end{equation}
where $H^{-1}$ is the size of the cosmological  horizon. The
cosmological horizon is similar to the event horizon of the  black
hole. Accordingly the Gibbons-Hawking temperature is defined by
$T_{\rm GH}=H/2 \pi$ ~\cite{GH} and the area of the horizon is $A=
4\pi/H^2$. A role of the Gibbons-Hawking temperature in inflation
was discussed in~\cite{Linde}. In order to see a route of
information flow from inflation to observable anisotropy, see the
Penrose diagram in Fig.1. \vspace{10mm}

\begin{figure}[here]
\hfil\scalebox{0.95}{\includegraphics{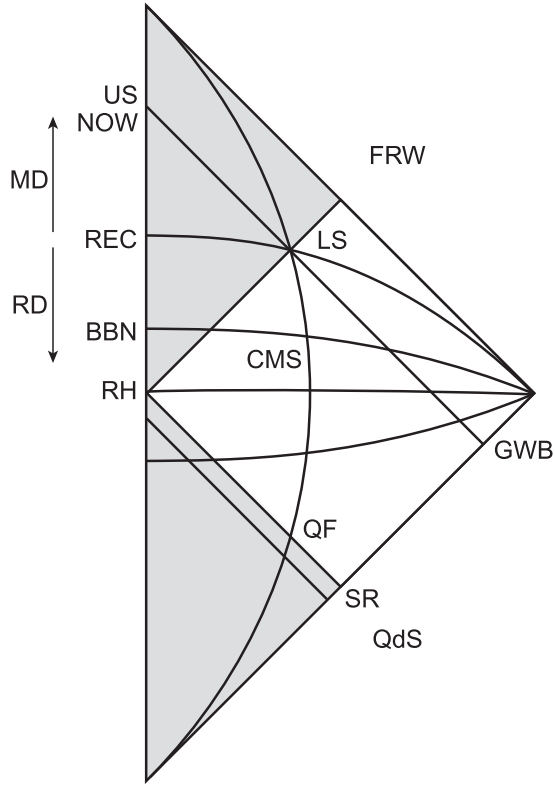}}\hfil
\caption{Penrose diagram of a cosmology with  inflation based on
quasi-de Sitter space and Friedmann-Robertson-Walker space.}
\end{figure}
\vspace{10mm} As usual, points denote two-spheres ($S^2$), the
left-hand edge represents the world line of an observer at the
origin. Others are boundaries at infinity. The lower half stands
for a quasi-de Sitter space (QdS) and the upper half for  a FRW
spacetime. The join between them is the epoch of reheating (RH)
and shaded regions of each show the regions within the apparent
horizon of an observer at the origin : one is an apparent
cosmological  horizon for QdS and
 the other is an apparent particle horizon for FRW.
 REC and BBN represent spacelike hypersurfaces for the
recombination epoch and big bang nucleosynthesis epoch. Two
regions in QdS are necessary, one appropriate for matching onto
FRW and the other for holographic analysis. Actually perturbations
may be  imprinted by fluctuating quantum field (QF) on the scale
of the apparent cosmological  horizon  during the slow-roll period
of inflation (SR). The apparent horizon grows slightly during SR,
as is shown by two closely parallel null lines. This happens
because the spacetime becomes asymptotically de Sitter space due
to the increase  of entropy during SR. The intersection of our
past light cone (null line) with REC is the two-sphere of the last
scattering surface (LS) for the cosmic background radiation. A
particular timelike trajectory of a comoving
 sphere (CMS) is shown. The radiation-dominated era (RD) is from
 the end of RH to the time of REC and the matter-dominated era (MD) is
 extended from REC to the present: US, NOW. Finally a high frequency
 gravitational wave background (GWR) can reach US via direct null
 trajectories.

In order to describe the inflation,  we introduce a scalar field
($\phi \equiv \phi(\tau)$: inflaton). This gives us the energy
density and pressure~\cite{LL}
\begin{equation}
\label{3eq3}
\rho_{\phi}=\frac{\dot \phi^2}{2} +V(\phi),~~
p_{\phi}=\frac{\dot \phi^2}{2} -V(\phi).
\end{equation}
Note that although the scalar field acts as a matter, it does not
possess an exact equation of state like $p_{\phi}=\omega_{\phi}
\rho_{\phi}$. Assuming a spatially flat universe, we obtain
\begin{equation}
\label{3eq4}
 H^2= \frac{1}{3 M_p^2} \Big[ V(\phi) +\frac{\dot \phi^2}{2} \Big]
\end{equation}
and
\begin{equation}
 \label{3eq5}
 \ddot \phi+ 3 H \dot \phi + V^{\prime}(\phi)=0
 \end{equation}
 where  the Planck mass is given by $M_p=1/\sqrt{8 \pi
G_4}$ in the units of $c=\hbar=1$. $V^{\prime}(\phi)$ denotes the
differentiation with respect to its argument.  The first equation
is obtained from Eqs.(\ref{2eq2}) and (\ref{3eq3}), whereas the
second from the conservation law of $\dot \rho +3H(\rho+p)=0$.
Inflation occurs when the potential energy of the scalar is
dominant in Eq.(\ref{3eq4}). This situation is approximated by the
slow-roll period of inflation which is formally defined by
\begin{equation}
\label{3eq6}
 \epsilon(\phi)=\frac{M^2_p}{2}\Big(
 \frac{V^{\prime}(\phi)}{V(\phi)} \Big)^2\ll1,~~
 |\eta(\phi)|\ll1~~{\rm with}~\eta=M^2_p \frac{
 V^{\prime\prime}(\phi)}{V(\phi)}.
 \end{equation}
 Then we obtain two equations (\ref{3eq4}) and (\ref{3eq5}) in the slow-roll approximation
\begin{equation}
\label{3eq7}  H^2 \simeq \frac{V(\phi)}{3M_p^2},~~
 3 H \dot \phi \simeq -V^{\prime}(\phi),
 \end{equation}
  where $\simeq$ indicates that the quantities are equal with the
 slow-roll approximation.
From Eq.(\ref{2eq2}) one finds a relation
\begin{equation}
\label{3eq9}  \dot H=-\frac{\dot \phi^2}{2M^2_p}.
\end{equation}
In the slow-roll approximation the potential can be taken to be a
nearly constant. Hence this can be approximated  by a quasi-de
Sitter phase  with its temperature $T_{\rm GH}$.

The slow-roll approximation is sufficient condition for inflation.
To see this, let us rewrite the condition for inflation as
\begin{equation}
\label{3eq9} \frac{\ddot R}{R}= \dot H +H^2 >0.
\end{equation}
This is obviously satisfied if $\dot H$ is positive. Otherwise, we
require
\begin{equation}
\label{3eq10}  \epsilon_{\rm HJ} \equiv- \frac{\dot H}{H^2}<1
\end{equation}
where $\epsilon_{\rm HJ}$ is the slow-roll parameter in the
Hamilton-Jacobi formalism. $\epsilon_{\rm HJ}$ leads to
\begin{equation}
\label{3eq11}
 \epsilon_{\rm HJ} \simeq \frac{3}{2} \frac{\dot \phi^2}{V(\phi)} \simeq
  \frac{M^2_p}{2}\Big(
 \frac{V^{\prime}(\phi)}{V(\phi)} \Big)^2 =\epsilon.
\end{equation}
Hence, if the slow-roll approximation is valid ($\epsilon \ll 1$),
then inflation ($\epsilon_{\rm HJ}<1$) is guaranteed.

On the other hand, when expressing $\epsilon_{\rm HJ}$ in terms of
he Hubble temperature $T_{\rm H}=-\frac{\dot H}{2\pi H}$ and the
Gibbons-Hawking temperature $T_{\rm GH}=\frac{H}{2 \pi}$, one
finds the relation
\begin{equation}
\label{3eq12} \epsilon_{\rm HJ}= \frac{T_{\rm H}}{T_{\rm
GH}}\simeq \epsilon.
\end{equation}
Similarly, if the slow-roll approximation is valid ($\epsilon \ll
1$),  an  inequality of $T_{\rm GH} > T_{\rm H}$ is guaranteed.
From Eq.(\ref{3eq10}) this inequality is another representation
for existing inflation.  Here the matter-temperature $T$ is
replaced by $T_{\rm GH}$ because in the slow-roll period of
quasi-de Sitter space, the matter-distribution is approximately
given by the positive cosmological constant ($\Lambda \simeq
V(\phi)$).
 From the above, we arrive at  the holographic temperature bound
\begin{equation}
\label{3eq13} T_{\rm GH} \ge T_{\rm H}
\end{equation}
which is valid for the period of inflation. This is the main
result of our work.  Eq.(\ref{3eq13}) has the same form as in the
Hubble temperature bound ($T \ge T_{\rm H}$) for a
radiation-dominated universe except replacing $T$ by $T_{\rm GH}$.
Here $T_{\rm H}$ corresponds to the minimum temperature in the
period of inflation. When the Gibbons-Hawking temperature is equal
to the Hubble temperature ($T_{\rm GH}=T_{\rm H}$), inflation
comes to an end ($\epsilon=1$).

Although two temperature of $T_{\rm GH}$ and $T_{\rm H}$ have the
geometric origin, our new interpretation for the inflationary
period are helpful to understand when the inflation ended. That
is, when $ T_{\rm GH}= T_{\rm H}$, inflation ended. However,
concerning the question of when the inflation started~\cite{BGV},
we have still no information.

\section*{Acknowledgment}
This work  was supported in part by KOSEF, Project Number: R02-2002-000-00028-0.

\end{document}